\documentclass[preprint,pre,oneside,floatfix,superscriptaddress,endfloats*]{revtex4}

\usepackage{bm}
\usepackage{amsmath}
\usepackage{graphicx}
\usepackage{tikz}
\usepackage[english]{babel}
\usepackage[dayofweek]{datetime}
\usepackage{color}

\newcommand{\dd}{\ensuremath{\mathop{}\!\mathrm{d}}}

\newcommand{\bs}[1]{\boldsymbol{#1}}

\definecolor{newalert}{RGB}{255, 10, 10}
\definecolor{newtext}{HTML}{0022CA}
\definecolor{oldnewtext}{RGB}{0, 104, 28}
\definecolor{orange}{rgb}{1,0.5,0}

\begin{document}


\title{Ferronematics in confinement}

\author{Grigorii Zarubin}
\email{zarubin@is.mpg.de}
\affiliation
{
   Max-Planck-Institut f\"ur Intelligente Systeme, 
   Heisenbergstr.\ 3,
   70569 Stuttgart,
   Germany
}
\affiliation
{
   Institut f\"ur Theoretische Physik IV,
   Universit\"at Stuttgart,
   Pfaffenwaldring 57,
   70569 Stuttgart,
   Germany
}
\author{Markus Bier}
\email{bier@is.mpg.de}
\affiliation
{
   Max-Planck-Institut f\"ur Intelligente Systeme, 
   Heisenbergstr.\ 3,
   70569 Stuttgart,
   Germany
}
\affiliation
{
   Institut f\"ur Theoretische Physik IV,
   Universit\"at Stuttgart,
   Pfaffenwaldring 57,
   70569 Stuttgart,
   Germany
}
\affiliation
{
   Fakult\"at Angewandte Natur- und Geisteswissenschaften,
   Hochschule f\"ur angewandte Wissenschaften W\"urzburg-Schweinfurt,
   Ignaz-Sch\"on-Str.\ 11,
   97421 Schweinfurt,
   Germany
}
\author{S.\ Dietrich}
\email{dietrich@is.mpg.de}
\affiliation
{
   Max-Planck-Institut f\"ur Intelligente Systeme, 
   Heisenbergstr.\ 3,
   70569 Stuttgart,
   Germany
}
\affiliation
{
   Institut f\"ur Theoretische Physik IV,
   Universit\"at Stuttgart,
   Pfaffenwaldring 57,
   70569 Stuttgart,
   Germany
}

\date{5 September 2018}

\begin{abstract}
The behavior of a uniformly magnetized ferronematic slab is investigated numerically
in a situation in which an external magnetic field is applied parallel and antiparallel, respectively, to its
initial magnetization direction. 
The employed numerical method allows one to determine hysteresis curves, from which a
critical magnetic field strength (i.e., the one at which the ferronematic sample becomes distorted) as function of the system parameters can be inferred.
Two possible mechanisms of switching the magnetization by applying a magnetic field in the
antiparallel direction are observed and characterized in terms of the coupling constant between
the magnetization and the nematic director as well as in terms of the coupling strength of the
nematic liquid crystal and the walls of the slab.
Suitably prepared walls allow one to combine both switching mechanisms in one 
setup, such that one can construct a cell the magnetization of which can be reversibly switched off.
\end{abstract}

\maketitle


\section{\label{sec:Intro}Introduction}

Ferronematics, i.e., suspensions of anisotropic ferromagnetic particles dispersed in a
nematic liquid crystal (NLC), attract both theoretical \cite{2006_Zadorozhnii, 2007_Zadorozhnii, 2014_Brand, 
2014_Zakhlevnykh, 2015_Zakhlevnykh, 2016_Boychuk, 2016_Zakhlevnykh, 2017_Zakhlevnykh} and experimental 
\cite{2011_Podoliak, 2013_Mertelj, 2014_Mertelj, 2015_Hess, 2017_Mertelj, 2018_Maximean} 
interest due to their ability to exhibit fluidity due to the solvent as well as macroscopic 
magnetization due to the colloidal inclusions. 
The anisotropic nature of the solvent implies broken rotational symmetry as compared to a
simple isotropic liquid. 
The interaction of the anisotropic ferromagnetic colloids with the solvent depends on the
orientation of the former with respect to the nematic director of the latter.
As a result, the individual magnetic moments of the colloids become effectively trapped 
around the two possible orientations of the nematic order. 
Therefore, suitably prepared samples can exhibit a macroscopically ferromagnetic phase. 
The phase behavior of this complex system follows from its free
energy density. 
The authors of Ref.~\cite{2013_Mertelj} proposed a phenomenological expression thereof which is formulated in terms of the magnetization
$\mathbf{M}$ and the nematic director $\mathbf{n}$. 
A similar expression was derived analytically starting from a microscopic description of
the system \cite{2018_Zarubin}:
\begin{equation}
  \label{eq:FN}
  f(\mathbf{M}, \mathbf{n})=
  \frac{a}{2}|\mathbf{M}|^2 - \frac{1}{2}\gamma\mu_0(\mathbf{M}\cdot\mathbf{n})^2
  - \mathbf{M}\cdot\mathbf{B}
\end{equation}
where $\mu_0=4\pi\times 10^{-7}$ N/A$^2$ is the permeability of vacuum, $a>0$ is a constant which depends on properties of both the nematic medium and the colloids (for an explicit form see Ref.\ \cite{2018_Zarubin}), $\gamma \geq 0$ measures the coupling between the magnetization and the nematic director, and $\mathbf{B}=B\mathbf{e}_x$ is the external magnetic field. 
Both $a$ and $\gamma$ are functions of the microscopic coupling constant $c:=WR/K$ where $W$
is the anchoring strength measuring the interaction energy of the NLC per 
surface area of a single colloid, $R$ is the radius of a colloidal particle modeled as a thin disc (i.e., a disc whose thickness is much smaller than its radius such that the interaction of the rim with the NLC medium can be disregarded), and $K$ is the elastic constant of the NLC within the one-elastic-constant approximation. (The value of $K=3.5\times10^{-12}$ N, corresponding to the twist elastic constant of 5CB \cite{2014_Mertelj}, is used throughout the current study if not specified otherwise.) 

One of the interesting results of the experiments reported in Ref.~\cite{2013_Mertelj} was the
observation of a complex response of the ferronematic slab to a uniform external magnetic
field, which depends on the initial state of the sample:
If, on one hand, the sample was prepared by quenching the NLC solvent from the 
isotropic into the nematic phase in the absence of an external magnetic field, the colloids formed \textit{various} magnetic domains within which $\mathbf{M}||\mathbf{n}$.
If, on the other hand, the NLC solvent was quenched in the presence of a uniform magnetic
field, a \textit{single} domain formed with the entire sample being magnetized in one 
direction with $\mathbf{M}||\mathbf{n}$. 
Applying thereafter a uniform external magnetic field opposite to the direction of the magnetization
of the single-domain sample yielded a complex, optically observable response of a 
nonuniform director field.

Here we focus on the case of single-domain samples. 
So far such samples have been thoroughly investigated theoretically in the situation in which the external magnetic field is applied \textit{perpendicular} to the initial magnetization of the sample \cite{2006_Zadorozhnii,2007_Zadorozhnii,2015_Zakhlevnykh}. 
Also the dynamics of such a configuration was investigated experimentally \cite{2017_Potisk,2018_Potisk,2018_Sebastian}. 
A \textit{thresholdless} distortion of the nematic was observed.
Moreover, the authors of Refs.\ \cite{2006_Zadorozhnii,2007_Zadorozhnii,2015_Zakhlevnykh} considered infinitely strong anchoring at the walls and external magnetic fields up to magnitudes strong enough to directly interact with the magnetically anisotropic NLC molecules.

Our aim is to investigate in detail the behavior of a monodomain sample exposed to a
uniform external magnetic field which is applied in the direction \textit{antiparallel} to the initial magnetization.
Moreover, we consider only magnetic fields of small ($\leq 25$ mT) amplitudes, such that the direct magnetic field influence on the NLC can be neglected. 
It was observed experimentally \cite{2013_Mertelj, 2014_Mertelj} that, like for common
ferromagnets, ferronematics exhibit hysteresis in the magnetic properties as a function
of the external field. 
Moreover, the critical field, i.e., the magnetic field strength at which the ferronematic becomes distorted (for a more precise definition see Sec.~\ref{subsec:I} below), is another feature of the ferronematic sample. 
We obtain the hysteresis curves numerically by using a conjugate-gradient technique in order to
minimize an appropriate free energy functional of the ferronematic in slab geometry. 
From the hysteresis curves one can infer the value of the critical magnetic field as 
function of the parameters of the model and compare them with the expressions derived in
Ref.~\cite{2013_Mertelj}. 
In our previous study \cite{2018_Zarubin} we derived the dependence of the coupling 
parameter $\gamma$ on the microscopic coupling $c$ which in turn depends on the particle 
size. 
Having obtained the critical field as a function of $\gamma$ allows us to relate it to the
size of the colloids and therefore one can potentially tune the value of the critical field
by tuning the mean of the size distribution of the particles used. 

It turns out that the switching process of the considered ferronematic slab from one phase to the other takes
place according to one of two possible scenarios which we shall discuss. 
In the first scenario, regions nucleate near the system walls in which the magnetization is flipped, whereas the nematic director is kept in place by the walls. 
In the second scenario, the nematic director follows the magnetization, i.e., it makes a full rotation by 180$^\circ$, everywhere throughout the sample.

Finally, we propose a novel technique which can be used, e.g., in data storage devices.
It is based on magnetic fields which control the magneto-optical properties of ferronematic
cells, and thus allows one to switch between magnetized and demagnetized states by applying a uniform magnetic field of suitable orientation.

The paper is organized as follows. 
In Sec.\ \ref{sec:Theory} we introduce the free energy functional in order to describe the system 
and the numerical method to minimize it. 
Section \ref{subsec:I} contains the description of the first of the two switching mechanism as well as
the results concerning the critical magnetic field and its dependence on the parameters of the 
model. 
In Sec.\ \ref{subsec:II} we present the second switching mechanism and provide a map which 
relates the parameters of the model to the character of the switching. 
In Sec.\ \ref{subsec:TwoWalls} we report that a combination of the two mechanisms leads to a sample the magnetization of which can be reversibly switched off by using the external magnetic field. 
The role of the phenomenon of segregation is discussed in Sec.~\ref{sec:Segregation}.
In Sec.\ \ref{sec:Discussion} we conclude by discussing the main results.


\section{\label{sec:Theory}Numerical model}

\begin{figure}[t!]
   \centering
   \includegraphics[scale=1]{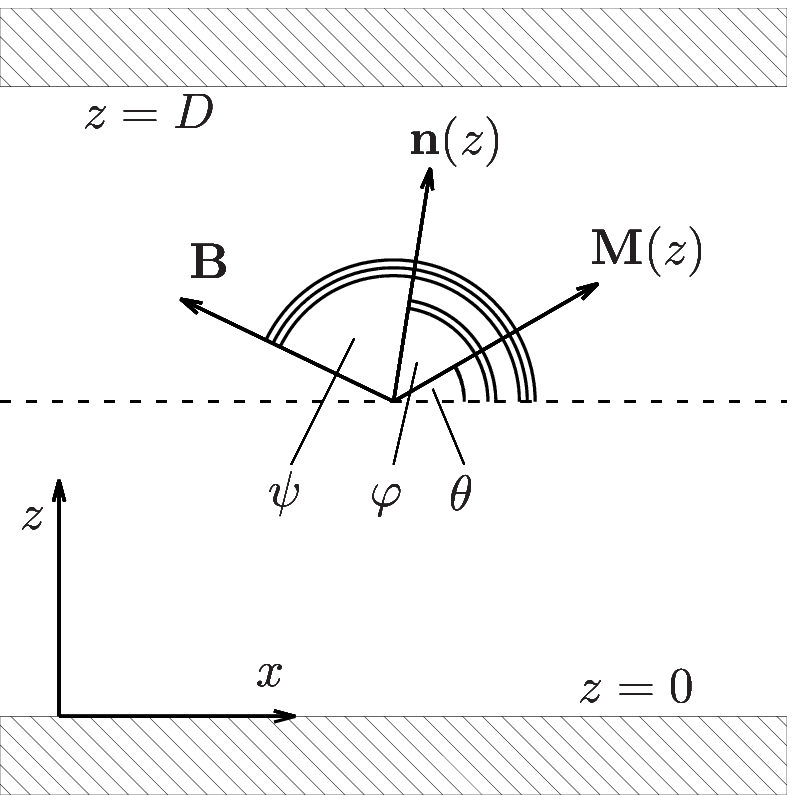}
   \caption{Sketch of a ferronematic in a slab of width $D$.
           The $x$-direction corresponds to the (lateral) easy direction of the liquid
           crystalline medium.
           Due to lateral translational invariance, all profiles depend only on the normal
           coordinate $z$.
           The nematic director $\mathbf{n}(z)$, the magnetization 
           $\mathbf{M}(z)$, and the external magnetic field $\mathbf{B}$ are parallel to the $x$-$z$-plane and their
           directions with respect to the positive $x$-direction are described by the angles $\varphi(z)$, $\theta(z)$, and $\psi$, respectively.}
   \label{fig:Slab}
\end{figure} 

We consider the experimental setup studied in Ref.~\cite{2013_Mertelj}. 
It consists of a ferronematic confined by two parallel and planar 
walls at a distance $D$ with $\mathbf{e}_x$ as the so-called easy
axis at both surfaces, which imposes a parallel orientation of the nematic 
director there (see Fig.~\ref{fig:Slab}). 
We assume that the sample was prepared in the presence of a homogeneous external magnetic
field $\mathbf{B}=B\mathbf{e}_x$ in the direction parallel to the easy axis $\mathbf{e}_x$ of the walls (i.e., $B>0$),
thus producing a single domain of the ferromagnetic phase. 
In the following the effect of applying an external magnetic field in the direction
opposite to the one used during this preparation (i.e., $B<0$) is investigated numerically.
The system is described by two spatially varying fields: the director field 
$\mathbf{n}(\mathbf{r})$ and the magnetization field $\mathbf{M}(\mathbf{r})$. 
We assume the absolute value of the magnetization is spatially constant,
$|\mathbf{M}(\mathbf{r})|=\text{const}=m\rho_\text{iso}$ (where $m$ is the absolute value of the magnetic moment of a single colloid, here taken to be $3\times 10^{-18}$~A~m$^2$ \cite{2013_Mertelj}, and $\rho_\text{iso}$ is the number density of the colloids dispersed in the isotropic phase of the liquid crystal during the preparation of the sample (see Refs.\ \cite{2013_Mertelj} and \cite{2018_Zarubin}), i.e., segregation effects are
assumed to be small \cite{footnote1}. 
(For a discussion of the possible influence of segregation see Sec.\ \ref{sec:Segregation}.)
Due to translational invariance in the lateral $x$-$y$-plane all physical quantities depend on the normal coordinate $z$ only. 
We consider that both $\mathbf{n}(z)$ and $\mathbf{M}(z)/(m\rho_\text{iso})$ are parallel to
the $x$-$z$-plane
\cite{footnote2}, 
so that they can be described by the angles $\varphi(z)$ and $\theta(z)$, respectively
(see Fig.~\ref{fig:Slab}). 
The initial configuration is given by the uniform profiles $\varphi(z)=0$ and 
$\theta(z)=0$, which corresponds to an unstable state when a uniform magnetic field 
$\mathbf{B}$ is applied in the direction $\psi=\pi$ (see Fig.~\ref{fig:Slab}).
In terms of the profiles $\varphi$ and $\theta$ the free energy functional of the system 
is given by 
\begin{equation}
\label{eq:SlabF}
  \frac{1}{S}\beta\mathcal{F}[\varphi,\theta] =
  \beta F_\text{ferr}[\varphi,\theta] + \beta F_\text{elas}[\varphi] + 
  \beta F_\text{surf}[\varphi],
\end{equation}
where $S$ is the surface area of one of the glass plates, $\beta:=1/(k_\text{B}T)$,
\begin{equation}
  \label{eq:Fferr}
   \beta F_\text{ferr}[\varphi,\theta]=
   \int_0^D\!\!\dd z\ \beta f(\mathbf{M}(z),\mathbf{n}(z))
\end{equation}
with the free energy density $f$ given by Eq.~\eqref{eq:FN}, which is the contribution due to the 
\textit{ferr}onematic,
\begin{equation}
  \label{eq:NLCEl}
  \beta F_\text{elas}[\varphi]=
  \frac{1}{2}\beta K\int_0^D\!\!\dd z\ \left( \frac{\dd\varphi(z)}{\dd z}\right)^2,
\end{equation}
is the contribution due to the elastic distortions of the liquid crystal, and
\begin{equation}
  \label{eq:WallC}
  \beta F_\text{surf}[\varphi]=
  -\frac{1}{2}\beta W_\text{wall}\left(\cos(\varphi(0))^2+\cos(\varphi(D))^2\right)
\end{equation}
is the contribution due to the coupling of the liquid crystal to the glass plates.

The equilibrium profiles $\varphi(z)$ and $\theta(z)$ correspond to the minimum of the free 
energy in Eq.~\eqref{eq:SlabF}, which has been determined numerically by using the 
Fletcher-Reeves-Polak-Ribiere general function minimization algorithm \cite{2002_Press}. 
The absolute value $|\mathbf{M}|$ of the magnetization is assumed to have a constant value $m\rho_\text{iso}$ and is taken to be independent of the external field $\mathbf{B}$ throughout Sec.\ \ref{sec:Results}. 
For the discussion of the problem in the case of a spatially varying $|\mathbf{M}|$ see Sec.\ \ref{sec:Segregation}.

We have used the following parameter values: $K=3.5\times 10^{-12}$~N, $\rho_\text{iso}=1.5\times 10^{19}$~m$^{-3}$, $m=3\times 10^{-18}$~A~m$^2$, and $T=300$~K. 
These values are consistent with experimental data reported in Refs.\ \cite{2013_Mertelj,2014_Mertelj}. 
The thickness $D$ of the slab is taken to be 20~$\mu$m throughout Sec.\ \ref{sec:Results}. 
Thicker slabs are considered in Sec.\ \ref{sec:Segregation} \cite{footnote7,1995_Burylov}.


\section{\label{sec:Results}Results}

\subsection{\label{subsec:I}Switching mechanism I and the critical field}

The experiments in Refs.~\cite{2013_Mertelj, 2014_Mertelj} demonstrate that, upon
applying a uniform external magnetic field to the setup described in Sec.~\ref{sec:Theory}
(see also Fig.~\ref{fig:Slab}), there is a nonvanishing critical magnetic field strength
$\mathbf{B}_\text{cr}=B_\text{cr}\mathbf{e}_x$ such, that for $B<B_\text{cr}<0$ ($B>B_\text{cr}>0$) in the case of an initial magnetization pointing along the positive (negative) $x$-direction, elastic distortions of the liquid
crystal matrix occur. 
The occurrence of such a critical magnetic field strength $B_\text{cr}$ can be 
explained qualitatively in terms of a diverging relaxation time of the fluctuations of the nematic
director field $\mathbf{n}$ (or $\varphi$) (see Ref.~\cite{2013_Mertelj}). 
Here we aim at exploring the dependence of $B_\text{cr}$ on the coupling constant 
$\gamma$ and the wall anchoring strength $W_\text{wall}$ \cite{footnote5}.
Moreover, we are also interested in the intermediate metastable states preceding the 
switches of the magnetization field $\mathbf{M}$ to the ground state parallel to the external field $\mathbf{B}$. 

\begin{figure}[t]
   \centering
   \includegraphics[scale=1]{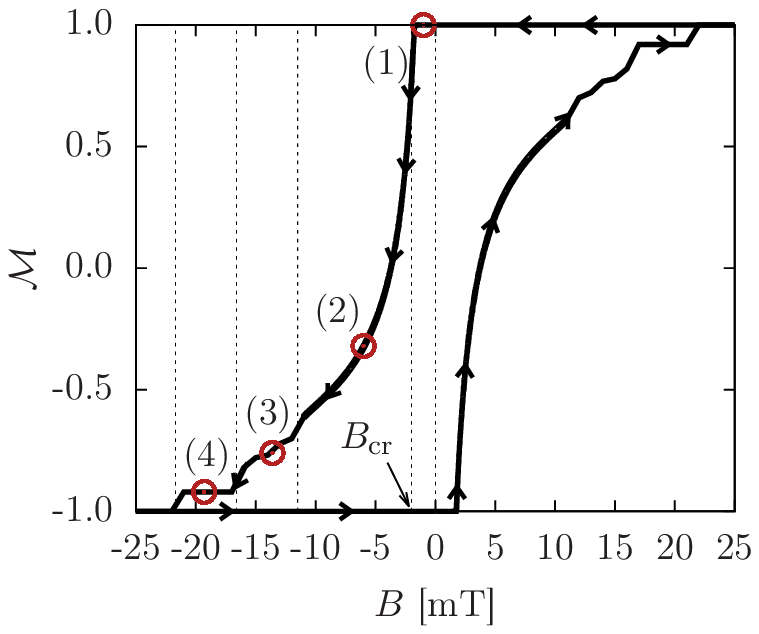}
   \caption{Hysteresis of the spatially averaged 
           magnetization $\mathcal{M}$ along the $x$-direction
           (see Eq.~\eqref{eq:PrettyM}) of the ferronematic slab as function of the 
           $x$-component $B$ of the external magnetic field $\mathbf{B}$. 
           Negative values of $\mathcal{M}$ or $B$ correspond to orientations in the 
           negative $x$-direction (see Fig.~\ref{fig:Slab}). 
           For initially saturated samples with $\mathcal{M}=1$ (i.e., magnetized in the positive $x$-direction)
           there is a nonvanishing critical magnetic field $B_\text{cr}<0$ 
           (indicated in the plot) such that for 
           $B\in[B_\text{cr},0]$ the magnetization $\mathcal{M}$ does not respond to the 
           external field.
           Upon increasing the field in the negative $x$-direction (i.e., for $B<B_\text{cr}$, 
             left branch of the loop) the 
             system evolves through a series of qualitatively distinct metastable states (red circles) 
             corresponding to the profiles displayed in Fig.~\ref{fig:MetaS} and eventually 
             reaches saturation along the negative $x$-direction (i.e., $\mathcal{M}=-1$). 
           Gradually lowering the magnitude $|B|$ of the magnetic field does not influence the 
           magnetization of the sample (i.e., for $B<0$, see the part of the loop along 
           $\mathcal{M}=-1$). 
           After $B=0$ is crossed, the situation is identical to the one 
           described above up to a change of sign of $B$ and $\mathcal{M}$ 
           (right branch of the loop).
           Dotted vertical lines separate regions of qualitatively different metastable
           states.
           Note that state (1) corresponds to $B_\text{cr}<B<0$.
           $W_\text{wall}=3.1\times 10^{-5}$~J/m$^2$, $\gamma=240$, $K=3.5\times 10^{-12}$~N, $\rho_\text{iso}=1.5\times 10^{19}$~m$^{-3}$, $m=3\times 10^{-18}$~A~m$^2$, and $T=300$~K}
   \label{fig:Hyst}
\end{figure} 
\begin{figure*}[t!]
   \centering
   \includegraphics[scale=0.85]{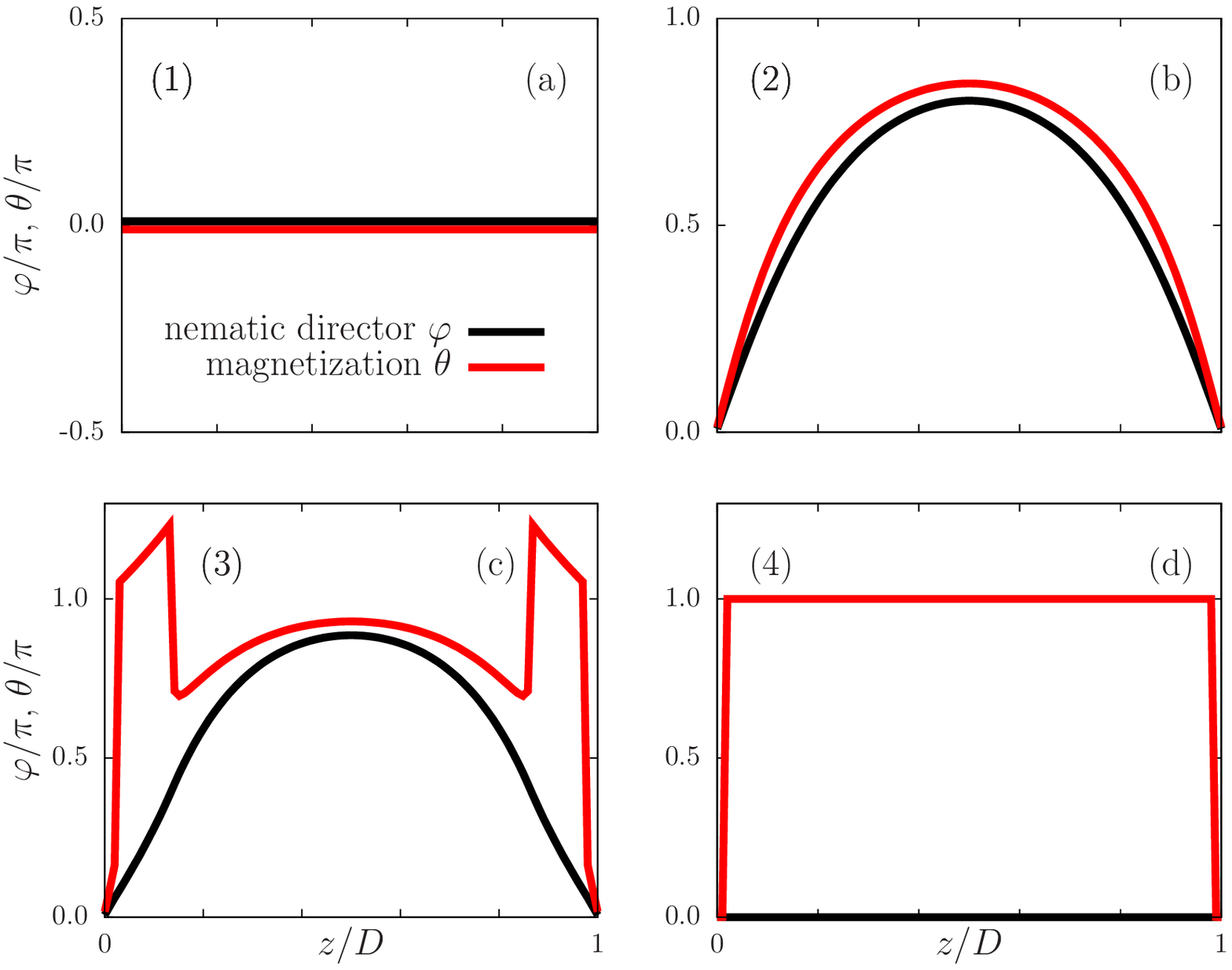}
   \caption{Qualitatively distinct metastable states of the ferronematic along the
           hysteresis curve in Fig.~\ref{fig:Hyst} in terms of the profiles 
           $\varphi(z)$ of the nematic director field (black lines) and $\theta(z)$
           of the local magnetization (red lines). 
           State (1) (see the numbered states in Fig.\ \ref{fig:Hyst}): 
           For magnetic fields $\mathrm{B}$ along the $x$-direction with
           components $B > B_\text{cr}$ the saturated profiles $\mathcal{M}=1$ are unperturbed by $B$. 
           State (2): For magnetic field components $B < B_\text{cr}$ both profiles deviate 
           significantly from the saturated ones.
           Note that the profiles in panel (1) correspond to a saddle point of the free energy
             so that spontaneous symmetry breaking can occur from state (1) to state (2) (i.e., 
             the magnetization and the director start to rotate in either clockwise or
             counterclockwise direction, see Fig.\ \ref{fig:Slab}).
           The equally probable profiles which correspond to the same projection $\mathcal{M}$ and thus correspond
           to the same points (1)-(4) on the hysteresis loop (see Fig.\ \ref{fig:Hyst}), are 
           obtained by the map $(\varphi,\theta)\mapsto(-\varphi,-\theta)$ for the profiles (a)-(d).
           State (3): Upon further increasing the magnetic field strength in the negative
           $x$-direction, layers form near the walls where the magnetization interpolates
           between the direction along ($\theta$ large) and opposite ($\theta$ small) to the magnetic field. 
           The coupling of the magnetization and the director causes $\theta>\pi$ within the layers.
           State (4): For even stronger magnetic fields the entire slab (besides thin
           layers near the walls, which require yet higher fields to switch) is magnetized
           along the direction of the magnetic field, i.e., $\theta=\pi$.
           The values of the system parameters are the same as in Fig.\ \ref{fig:Hyst}.}
   \label{fig:MetaS}
\end{figure*} 

Figure \ref{fig:Hyst} shows the hysteresis curve of the spatially averaged projection of the magnetization $\mathcal{M}$
onto the $x$-axis,
\begin{equation}
  \label{eq:PrettyM}
  \mathcal{M}:=\frac{1}{D}\int_0^D\!\!\dd z\,\,\cos\theta (z),
\end{equation}
as function of the component of the external magnetic field $\mathbf{B}$ along the $x$-axis
for the particular choice of the coupling constant $\gamma=240$ (which corresponds to a value of the microscopic coupling constant $c\approx 0.035$ \cite{2018_Zarubin}) and of the wall anchoring strength $W_\text{wall}=3.1\times 10^{-5}$ J/m$^2$; this choice of the parameters is reasonable in the context of available experimental data (see Refs.\ \cite{2013_Mertelj}, \cite{2014_Mertelj}, and \cite{2018_Zarubin}). 
In order to investigate the switching process of the magnetization in the ferromagnetic
phase as function of the external magnetic field, Fig.~\ref{fig:MetaS} displays the
orientation profiles $\varphi(z)$ and $\theta(z)$ for a series of intermediate metastable
states corresponding to the hysteresis loop in Fig.~\ref{fig:Hyst}. 

For initially saturated samples with $\mathcal{M}=1$, the magnetization does not change
significantly in the presence of $x$-components of the magnetic field $B > B_\text{cr}$, whereas for 
$B < B_\text{cr}$ there is a noticable deviation of the $x$-component of the spatially averaged 
magnetization $\mathcal{M}$ from the initial saturation value (see Fig.~\ref{fig:Hyst}).
This defines a critical magnetic field strength $B_\text{cr} < 0$.
For $B > B_\text{cr}$ both the magnetization and the nematic director field profiles, 
i.e., $\theta(z)$ and $\varphi(z)$, de facto do not deviate from the saturated ones (see Fig.~\ref{fig:MetaS}(a)).
While the magnetization tends to align with the external magnetic field, due to the interaction described by the coupling constant $\gamma$ it is dragging the nematic director field along. 
At $B=B_\text{cr}$ the metastable state corresponding to the unperturbed nematic director
becomes unfavorable compared to the metastable state corresponding to the perturbation
induced in the interior of the slab (see Fig.~\ref{fig:MetaS}(b)). 
We note that the saturated sample with $\mathcal{M}=1$ in a magnetic field in the negative
$x$-direction with $B=B_\text{cr}$ corresponds to a saddle point of the free energy so
that spontaneous symmetry breaking induced by fluctuations leads to perturbations
of the magnetization orientation profile $\theta(z)$ with either $\theta(z) > 0$ or
$\theta(z)<0$. 
In the following we focus only on the first case, while the second, conjugated one, follows from changing signs.
It is the perturbed nematic director field which manifests itself as a brightening of the
sample when viewed with crossed polarizers as in the experiment reported in Ref.~\cite{2013_Mertelj},
and it occurs only due to the coupling of the magnetization field $\mathbf{M}$ to the 
nematic director field $\mathbf{n}$. 
The external magnetic field imposes a torque onto the magnetization field which in turn leads to a 
torque onto the nematic director field. 
The latter is opposing the torque generated by the walls
of the cell and which is transmitted due to the elasticity of the NLC (Eq.~\eqref{eq:NLCEl}). 
Upon increasing the external magnetic field the variations inside the slab become more
and more pronounced for both the magnetization and the nematic director field. 
However, in the case of soft anchoring \cite{footnote6} (distinct from the case of infinitely strong anchoring, see Ref.~\cite{1970_Brochard}) at
the surface of the colloid, the angle between the 
magnetization and the nematic director is nonzero for $B < B_\text{cr}$ (see 
Figs.~\ref{fig:Slab} and \ref{fig:MetaS}), i.e., $\theta\neq\varphi$.

\begin{figure}[t!]
   \centering
   \includegraphics[scale=1]{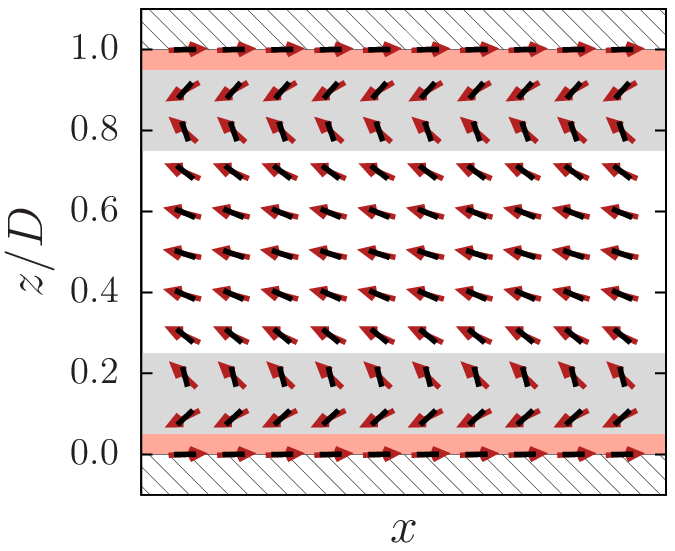}
   \caption{Explicit magnetization field (red arrows) and nematic director field (black rods) of the pretransitional metastable state (3) (see 
           Figs.~\ref{fig:Hyst} and \ref{fig:MetaS}(c)) of the ferronematic in between two
           glass walls (hatched regions). 
           While in the interior of the slab (white region) the magnetization field (red arrows) is, to a large extent, aligned with the external magnetic
           field in the negative $x$-direction,
           close to the walls (grey regions) its orientation interpolates between the configurations being parallel and antiparallel to
           the magnetic field. 
           This is due to the coupling to the
           nematic director field (black rods), which is aligned along the easy direction at the
           walls. 
           Upon further increasing the strength of the external magnetic field in the negative 
           $x$-direction ($B < B_\text{cr} < 0$), the grey regions widen and
           eventually produce an almost uniformly magnetized sample (see 
           Fig.~\ref{fig:MetaS}(d)).
           The pale red regions denote thin layers very close to the walls which switch last.}
   \label{fig:Nucleation}
\end{figure}

Before reaching the magnetic phase with the sample being magnetized along the field in negative $x$-direction, the
system passes through the metastable state (3) in Fig.~\ref{fig:Hyst}, in which the 
magnetization profile $\theta(z)$ has a peculiar form (see Figs.~\ref{fig:MetaS}(c) and 
\ref{fig:Nucleation}). 
In this metastable state the magnetization in the interior of the sample is aligned along
the magnetic field.
Within certain transition regions close to the walls the orientation of the magnetization interpolates between the direction along the
magnetic field and the opposite direction.
These transition regions occur because the magnetization is coupled to the nematic director
field, which is aligned along the easy axis ($\varphi =0$) at the walls.
The width of these transition regions grows upon increasing the external magnetic field strength so that
eventually the minimum of the free energy given by  Eq.~\eqref{eq:SlabF} corresponds to 
the magnetization being oriented parallel to the external magnetic field in the entire slab ($\theta=\pi$). 
Concerning the transition regions following observations can be made: 
(i)  Due to the soft coupling between the colloids and the nematic director field of the 
     NLC the ground state, in which the entire sample is magnetized along the external field,
     is attained by means of ``switching'' the magnetization locally, i.e., by inverting
     the direction of the magnetization (and thus of the orientation of the magnetic colloids)
     without the simultaneous rotation of the local nematic director field. 
(ii) Layers of the incipient ferronematic phase are nucleated in the regions close to the walls
     due to the interplay between elastic and magnetic torques and because
     the coupling energy is invariant with respect to an inversion of the magnetization.

\begin{figure}[t!]
   \centering
   \includegraphics[scale=1]{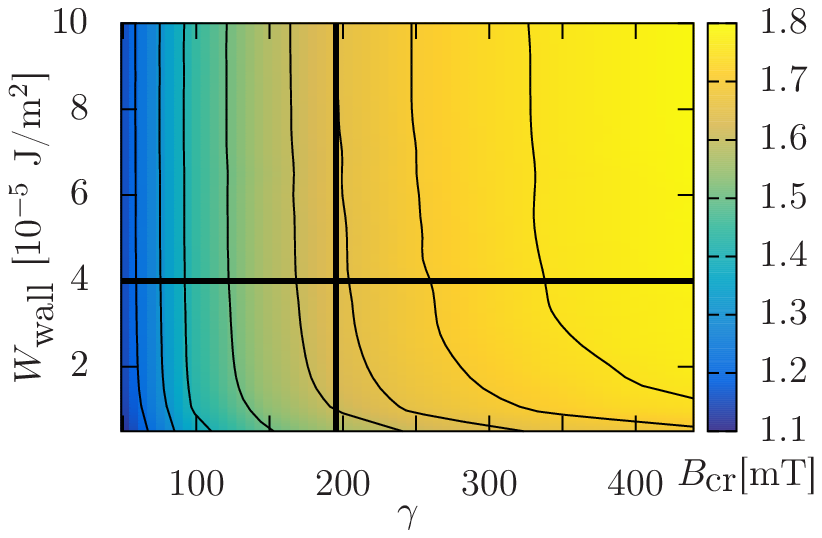}
   \caption{The dependence of the critical magnetic field strength $B_\text{cr}$ on the 
           coupling constant $\gamma$ and on the anchoring strength $W_\text{wall}$ at the
           walls.
           The color code denotes values of $B_\text{cr}$ measured in units of mT. 
           The thin black curves are contour lines; from left to right $B_\text{cr}=1.2,1.3,1.4,1.5,1.6,1.65,1.7,1.75$ mT.
           The thick black lines correspond to
           the cuts in Fig.~\ref{fig:Cuts}(a) and (b) for $\gamma\approx 195$ and 
           $W_\text{wall}=4\times 10^{-5}$ J/m$^2$, respectively (the values of the remaining parameters are the same as in Fig.\ \ref{fig:Hyst}).}
   \label{fig:Phase}
\end{figure}

\begin{figure}[t!]
   \centering
   \includegraphics[scale=1]{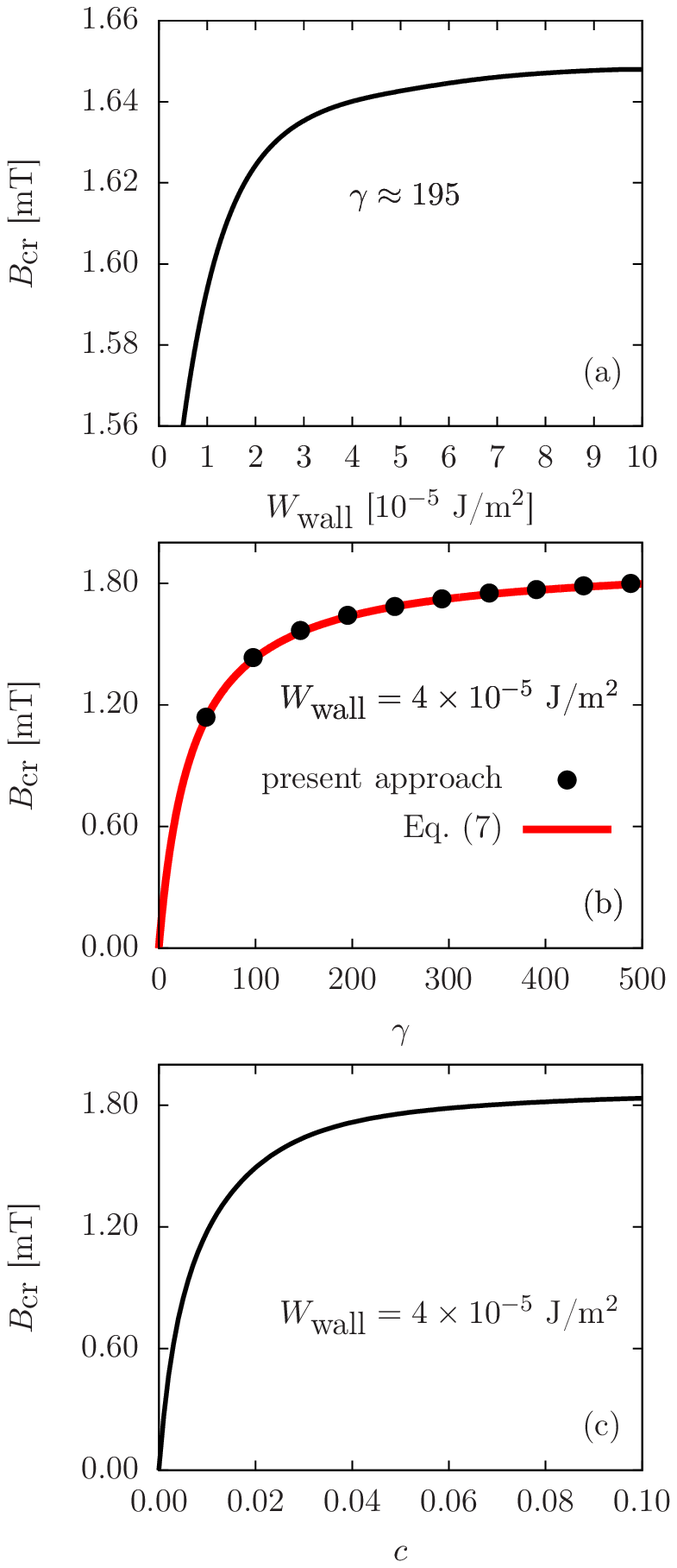}
   \caption{Dependence of the critical magnetic field strength $B_\text{cr}$
           (a) on the wall anchoring strength $W_\text{wall}$ (along the thick vertical black line in Fig.~\ref{fig:Phase}, i.e., for $\gamma\approx 195$), 
           (b) on the coupling constant $\gamma$ (present approach and Eq.\ \eqref{eq:BCr} along the thick horizontal black line in 
           Fig.~\ref{fig:Phase}, i.e., for $W_\text{wall}=4\times 10^{-5}$ J/m$^2$), and 
           (c) on the microscopic coupling constant $c$ (see see below Eq.\ \eqref{eq:FN}, $W_\text{wall}=4\times 10^{-5}$ J/m$^2$). The values of the remaining parameters are the same as in Fig.\ \ref{fig:Hyst}.}
   \label{fig:Cuts}
\end{figure}

Naturally the question arises concerning the dependences of the critical magnetic field strength
$B_\text{cr}$ on the coupling constant $\gamma$ and on the wall anchoring $W_\text{wall}$. 
Here we define $B_\text{cr}$ as the magnetic field strength at which the spatially averaged magnetization 
$\mathcal{M}$ equals $0.97$; note that $\mathcal{M}=1$ in the saturated state. 
This definition differs from the one used in Refs.~\cite{2013_Mertelj, 2014_Mertelj},
where $B_\text{cr}$ is defined as the magnetic field strength at which the relaxation time
of thermal fluctuations of the direction of $\mathbf{n}$ diverges. 
Here we do not consider dynamic processes, instead we propose the above alternative definition of
$B_\text{cr}$. 
Obviously, the choice of $0.97$ for the threshold value contains some degree of arbitrariness.
However, as can be inferred
from the steep slope of the hysteresis loop close to state (1) in Fig.~\ref{fig:Hyst}, no significant changes are
expected to occur by choosing different threshold values not too much less than unity. 
Figure \ref{fig:Phase} shows the dependence of $B_\text{cr}$ on the coupling constant $\gamma$ and
the wall anchoring strength $W_\text{wall}$. 
One can infer from Fig.~\ref{fig:Phase} that for fixed $W_\text{wall}$ the critical field 
$B_\text{cr}$ increases upon increasing $\gamma$. 
Indeed, for a given value of $W_\text{wall}$, the magnetization field is aligned with the
nematic director field, the rotation of which is opposed by the torque imposed by the
walls. 
The system sustains the alignment for increasing external magnetic field strengths which in their turn are due to an increasing strength of the coupling
$\gamma$ between the magnetization and the nematic director field. 
One can also infer from Fig.~\ref{fig:Phase} that the critical field $B_\text{cr}$ 
depends rather weakly on the wall anchoring strength $W_\text{wall}$: Within the considered range of the
anchoring strengths, $W_\text{wall}\in[0.5\times 10^{-5},10\times 10^{-5}]\mathrm{J/m^2}$, for fixed $\gamma$ the
critical field strength $B_\text{cr}$ varies by $\approx 0.1\text{ mT}$. 
For large values of $W_\text{wall}$ the critical field reaches a plateau (see Fig.\ \ref{fig:Cuts} (a))
and it becomes independent of the wall anchoring $W_\text{wall}$. 

Figure~\ref{fig:Cuts} illustrates these trends of the critical magnetic field strength
$B_\text{cr}$ via cuts in Fig.~\ref{fig:Phase} for $\gamma\approx 195$ (see 
Fig.~\ref{fig:Cuts}(a)) and $W_\text{wall}=4\times 10^{-5}$ J/m$^2$ (see Fig.~\ref{fig:Cuts}(b)).
Moreover, Fig.~\ref{fig:Cuts}(c) displays the dependence of $B_\text{cr}$ on the 
microscopic coupling constant $c$ (see below Eq.\ \eqref{eq:FN}).
Finally, Fig.~\ref{fig:Cuts}(b) compares the values of the critical magnetic field 
$B_\text{cr}$ as defined in the present approach (circles) with the corresponding expression given in Ref.~\cite{2013_Mertelj}, 
\begin{equation}
  \label{eq:BCr}
  B^\text{\cite{2013_Mertelj}}_\text{cr}=
  \frac{\pi^2 \gamma \mu_0 K M_\text{s}}{\pi^2 K+\gamma\mu_0 M_\text{s}^2D^2},
\end{equation}
where $M_\text{s}:=m\rho_\text{iso}$ is the magnetization of the saturated sample, obtained in the limit $W_\text{wall}\to\infty$. 
(Here, we consider the particular NLC used for the experiments in Refs.\ \cite{2013_Mertelj,2014_Mertelj} and therefore a fixed value of the elastic constant $K$. 
Although the variation of the expression given here as function of the elastic constant of the
NLC is interesting, we leave this issue for future work due to the highly non-trivial 
occurrences of $K$.)
Remarkably, $B_\text{cr} \approx B^\text{\cite{2013_Mertelj}}_\text{cr}$ appears to hold
although the two definitions of the critical magnetic field strength differ and although
$B^\text{\cite{2013_Mertelj}}_\text{cr}$ in Eq.~\eqref{eq:BCr} does not take the dependence
on $W_\text{wall}$ into account.


\subsection{\label{subsec:II}Switching mechanism II}

In Sec.~\ref{subsec:I} we revealed a mechanism of switching the sample magnetization in the case of large values of the wall anchoring strength $W_\text{wall}$. 
Here we show another possible mechanism which corresponds, however, to small values of
$W_\text{wall}$.
For suitable combinations of $\gamma$ and $W_\text{wall}$ the magnetization field is able to
drag the nematic director field along, thereby inducing a large change of the angle 
$\varphi$ compared with the initial configuration. 
For small values of $W_\text{wall}$ the anchoring at the wall is so weak, that the nematic director field at the surface of the walls is able to deviate from the direction of the easy axis and to rotate with the magnetic field due to the coupling between the magnetization and the nematic director. 
\begin{figure*}[t!]
   \centering
   \includegraphics[scale=0.85]{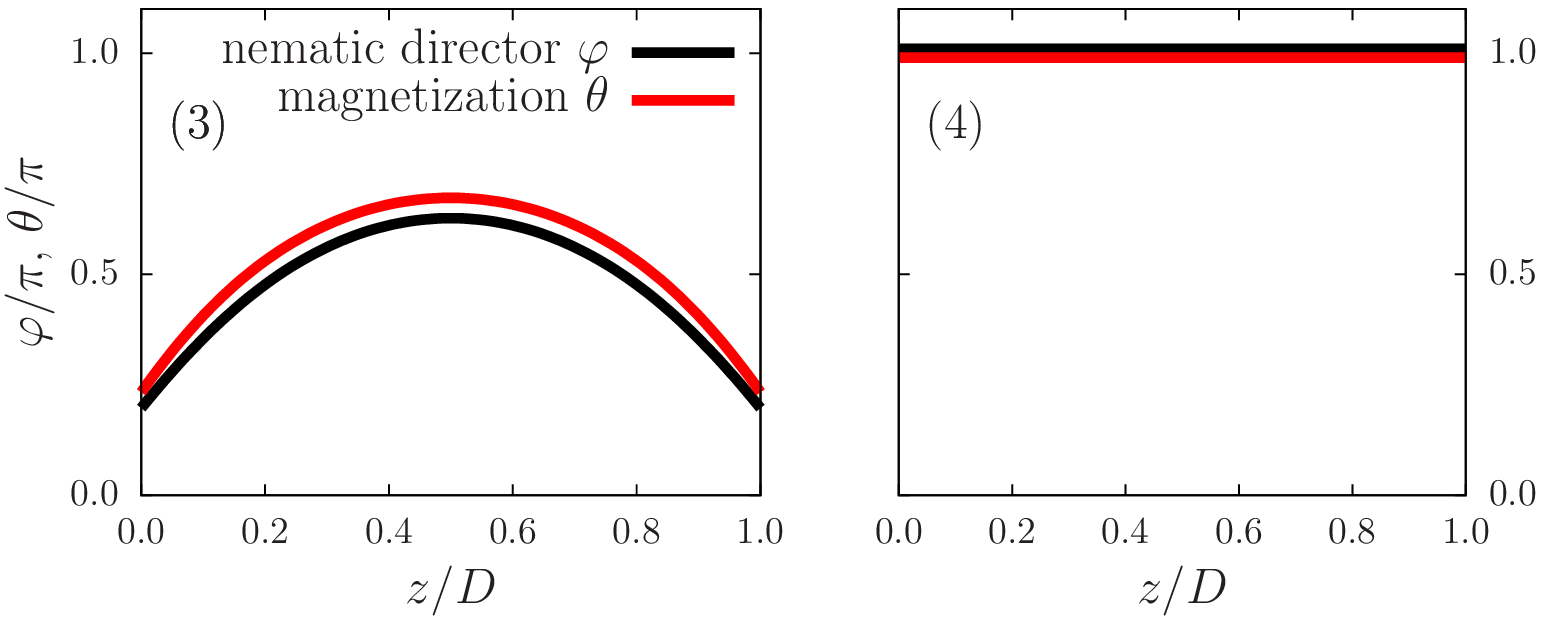}
   \caption{Final two stages of switching mechanism II for the magnetization in an initially
           oppositely oriented external magnetic field for $\gamma=240$ and $W_\text{wall}=0.1\times 10^{-5}$~J/m$^2$ (the values of the remaining parameters are the same as in Fig.\ \ref{fig:Hyst}).
           It is characterized by a nematic director field with significant elastic 
           distortions throughout the sample. 
           Even at the surface rotations of the nematic director with respect to the easy
           axis occur due to the weak anchoring at the walls (note $\varphi\neq 0$ and $\theta\neq 0$ in the left panel). 
           For sufficiently strong magnetic fields both the magnetization field and the 
           nematic director rotate by the angle $\pi$ (see right panel). 
           The panels are denoted by ``(3)'' and ``(4)'' in order to make the comparison easier
           with the corresponding panels in Fig.~\ref{fig:MetaS} describing switching mechanism I.
       Note that $\theta\neq\varphi$ even directly at the walls.}
   \label{fig:MetaSII}
\end{figure*} 

We have performed calculations analogous to those described in Sec.~\ref{subsec:I} but for 
small values of the anchoring strength $W_\text{wall}$ at the sample walls. 
It turns out that for values $W_\text{wall} <0.5\times 10^{-5}$~J/m$^2$ there are 
corresponding values of the coupling constant $\gamma$ which produce a switching mechanism which 
is qualitatively different from the one described in the previous section.
In this mechanism the early stages of the switching are similar to those described in the
previous section (see Fig.~\ref{fig:MetaS}(1) and (2)). 
However, the subsequent stage, as displayed in Fig.~\ref{fig:MetaSII}(3), is qualitatively different 
in the sense that the system does not separate in distinct spatial regions with 
different orientations of the magnetization (compare with Fig.~\ref{fig:MetaS}(3)).
Obviously, the change in character of the switching mechanism is directly related to the
anchoring at the walls being too weak to prevent the liquid crystal from rotating along 
with the magnetization field. 
This weakness is revealed also by nonzero angles $\varphi$ and $\theta$ at the walls.

\begin{figure}[t!]
   \centering
   \includegraphics[scale=1]{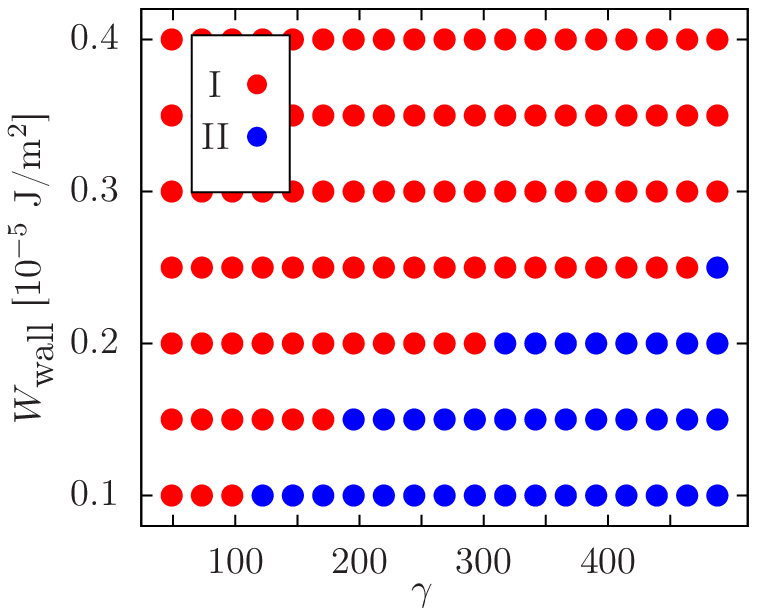}
   \caption{Regions of dominance for the two switching mechanisms between two ferromagnetic phases in
           terms of the coupling constant $\gamma$ and the wall anchoring strength $W_\text{wall}$, which is the same for both walls. (The values of the remaining parameters are the same as in Fig.\ \ref{fig:Hyst}.) 
           The region denoted as "I" (full red circles) corresponds to switching mechanism I in which the 
           magnetization leaves the nematic director field behind (see 
           Sec.~\ref{subsec:I}).
           The region denoted as "II" (full blue circles) corresponds to switching mechanism II in which the
           nematic director field is weakly coupled to the sample walls and therefore is
           able to follow the magnetization.}
   \label{fig:Switch}
\end{figure}

At such low anchoring strengths the torque imposed on the liquid crystal director by the 
walls cannot compete with the drag imposed by the rotating magnetization field and it is 
energetically more favorable for the director at the walls to flip its orientation. 
The regions of dominance for the two switching mechanisms are depicted in Fig.~\ref{fig:Switch}, where a map spanned by the
coordinates ($\gamma$, $W_\text{wall}=W_\text{wall}^{(1)}=W_\text{wall}^{(2)}$) marks region "I" (corresponding to switching mechanism I, for which
the director field returns back to its initial configuration upon increasing the external magnetic field and reaching the saturation of the magnetization in the direction of the field) and region "II" (corresponding to switching mechanism II, for which the 
director field follows the magnetization).
Within our numerical approach switching mechanism II is observed only for 
values of the wall anchoring $W_\text{wall}$ which are significantly smaller than the one
estimated from the experiment \cite{2014_Mertelj}, i.e., $W^\text{exp}_\text{wall}\approx 
(3.40\pm 0.11)\times10^{-5}$~J/m$^2$. 
Therefore, we expect switching mechanism I to be the one realized experimentally.

The segregation of colloids might play an important role.
Segregation amounts to a redistribution of the colloids dispersed in the liquid crystal. 
This effect is caused by the opportunity to lower the free energy of the magnetic colloids 
in an external magnetic field by migrating away from regions in which the liquid crystal
prevents alignment along the external field. 
Thereby energy is gained by accomplishing alignment at the expense of the entropic contribution due to denser packing which is proportional to $\rho\log\rho$, where 
$\rho$ is the local number density of the colloids (see Ref.~\cite{2018_Zarubin}). 
Although segregation is neglected in Sec.\ \ref{sec:Results}, we nonetheless do not expect segregation to influence our results qualitatively (see Sec.\ \ref{sec:Segregation}). 
On the other hand, quantitative changes are conceivable, i.e., the map in Fig.~\ref{fig:Switch} might be affected.


\subsection{\label{subsec:TwoWalls}Confining walls with different anchoring strengths}

In this section we study a combination of switching mechanisms I and II described 
in Secs.~\ref{subsec:I} and \ref{subsec:II}, respectively, by considering a strong anchoring
strength at one wall and a weak one at the other, sharing the same easy axis.
Thus the system is described by three parameters (assuming $a$, $T$, $K$, $D$, $m$, and $\rho_\text{iso}$ to be fixed): (i) coupling constant $\gamma$, (ii) the
anchoring strength at one of the walls, and (iii) the ratio of the anchoring strengths
at the two walls. 
Note that introducing different but still uniform and parallel anchorings at the walls renders the 
system still effectively one-dimensional along the $z$-direction and hence from a numerical point of view its complexity does not change.

Adding a third parameter (i.e., the ratio of the anchoring strengths at the two walls $w:=W_\text{wall}^{(1)}/W_\text{wall}^{(2)}$) introduces a third dimension to the map considered in Fig.~\ref{fig:Switch}. 
The cut of this three-dimensional map along $w=1$ produces the two-dimensional map shown in Fig.~\ref{fig:Switch}.
While the two-dimensional map in Fig.~\ref{fig:Switch} exhibits only two switching regions (I and II), in the three-dimensional parameter space the situation can be more involved. 
It is reasonable to expect that if both $W_\text{wall}^{(1)}$ and $W_\text{wall}^{(2)}$ become infinitely strong, the magnetization of the sample switches according to mechanism I. If, on the other hand, $W_\text{wall}^{(1)},W_\text{wall}^{(2)}\to 0$ one can expect that the magnetization of the sample switches according to mechanism II.
However, pairs $(W_\text{wall}^{(1)},W_\text{wall}^{(2)})$ can exist such that the magnetization in the vicinity of one wall would switch according to mechanism I and the magnetization in the vicinity of the other wall would switch according to mechanism II. 
Thus, the three-dimensional parameter space consists of \textit{three} regions: dominance of mechanism I, II, and their combination.

\begin{figure}[t!]
   \centering
   \includegraphics[scale=1]{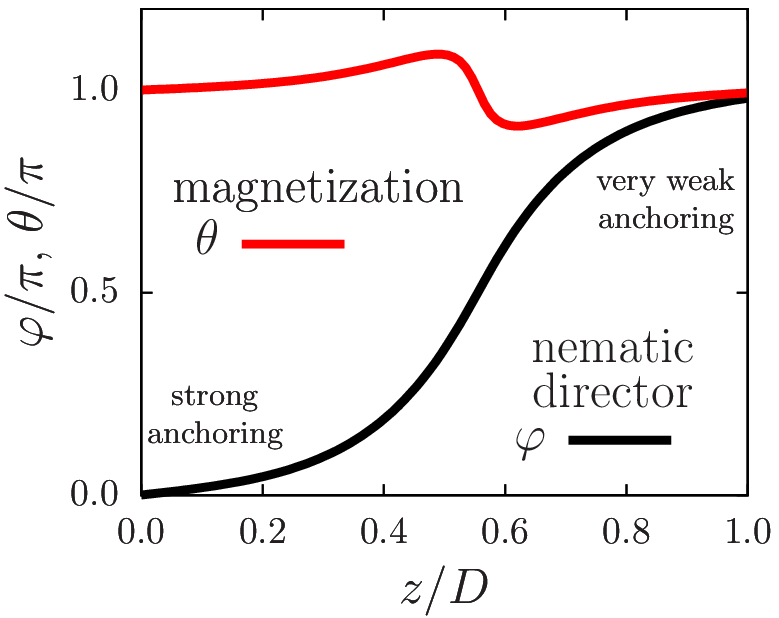}
   \caption{Orientation profiles $\varphi(z)$ (nematic director, black line) and $\theta(z)$ 
           (magnetization, red line) for $B \leq -12$ mT (with the sample initially magnetized along the positive $x$-direction).
           There is a gradual variation of the nematic director from one wall to the other.
           The parameters are chosen as $\gamma=240$, $W^{(1)}_\text{wall}=10^{-5}$~J/m$^2$ (see $z=0$), and 
             $W^{(2)}_\text{wall}=0.1\times10^{-5}$~J/m$^2$ (see $z=D$), $K=9\times 10^{-12}$ N \cite{footnote3}; the values of the remaining parameters are the same as in Fig.\ \ref{fig:Hyst}.}
   \label{fig:TwoWalls}
\end{figure}

The line separating the two regions in Fig.~\ref{fig:Switch} becomes a two-dimensional manifold in the three-dimensional parameter space ($W_\text{wall}^{(1)},w,\gamma$). 
In the vicinity of the plane $w=1$ this manifold, which is separating the regions ``I'' and ``II'', is considered to be perpendicular to the plane $w=1$ and only the two regions ``I'' and ``II'' occur. 
Therefore, if one would like to find a point ($W_\text{wall}^{(1)},w,\gamma$) that belongs to the region, which corresponds to the combination of the two switching mechanisms, it is necessary to pick the value of $w$ significantly different from 1. 
To this end, for fixed $\gamma$ it seems to be natural to take as an estimate the anchoring at one of the walls from region ``I'' in Fig.~\ref{fig:Switch} and the anchoring at the other wall from region ``II'' in Fig.~\ref{fig:Switch}.
It has turned out that for $D=20$~$\mu$m (for a discussion concerning larger values of $D$ see Sec.\ \ref{sec:Segregation}) the combination $W_\text{wall}^{(1)}=10^{-5}$~J/m$^2$ and  $W_\text{wall}^{(2)}=0.1\times 10^{-5}$~J/m$^2$ (i.e., $w=10$) yields profiles $\varphi(z)$ (nematic director) and $\theta(z)$ 
(magnetization) which consist of one part due to switching mechanism I and another 
part due to switching mechanism II. 
Figure \ref{fig:TwoWalls} shows the actual profiles (i.e., for magnetic field strengths 
$B \leq -12$ mT and for the initial magnetization pointing into the positive $x$-direction) of the magnetization and of the nematic
director field for $\gamma=240$, $W^{(1)}_\text{wall}=10^{-5}$ J/m$^2$, and
$W^{(2)}_\text{wall}=0.1\times10^{-5}$ J/m$^2$, where superscript $(1)$ denotes the wall at $z=0$ and superscript $(2)$ denotes the wall at $z=D$.
Both the orientation field $\varphi(z)$ of the nematic director and the orientational field $\theta(z)$ of the magnetization have a nontrivial form. 
The nematic director field profile exhibits a smooth rotation by an angle of $\pi$ from
one wall to the other. 
According to Fig.\ \ref{fig:TwoWalls}, the wall with the strong anchoring at $z/D=0$ is able to 
align the nematic director along the easy axis there ($\varphi=0$), while the magnetization field is 
switching to the negative $x$-direction parallel to the external field (compare Sec.~\ref{subsec:I}).
On the other side, the weak anchoring at $z/D=1$ in Fig.~\ref{fig:TwoWalls} allows the nematic director there 
to follow the magnetization ($\theta=\pi$ implies $\varphi=\pi$; compare Sec.~\ref{subsec:II}). 
This provides a situation in which $\varphi=0$ at one wall and $\varphi=\pi$ at the other. 
The elastic contribution in Eq.~\eqref{eq:NLCEl} ensures that no singularities occur in the
interior of the slab so that there is a smooth crossover between the two boundary values.

\begin{figure}[t!]
   \centering
   \includegraphics[scale=1]{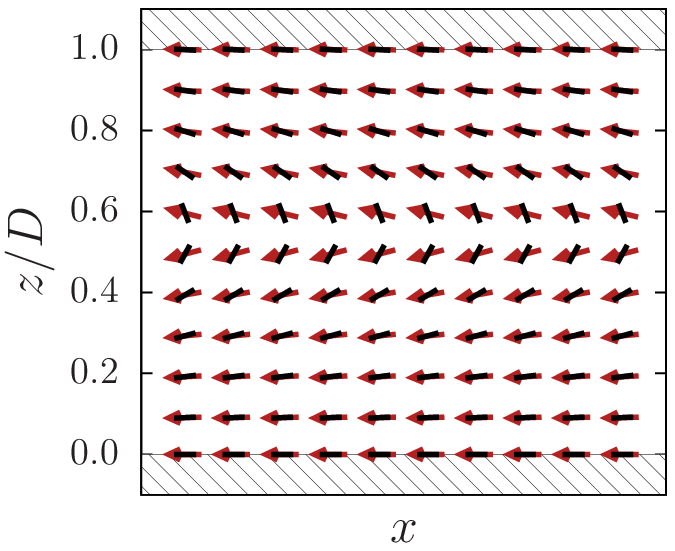}
   \caption{Vector fields corresponding to Fig.~\ref{fig:TwoWalls}. 
           The notation is the same as in Fig.~\ref{fig:Nucleation}. 
           Note the distortion of the nematic director field (black rods) 
           in the interior and the 
           interface between two magnetization (red arrows) domains at $z\approx 0.55D$ due to
           the combination of the switching mechanisms I and II. 
           Basically throughout the whole sample the magnetization has reached the switched state in negative $x$-direction.
         $\gamma=240$, $W^{(1)}_\text{wall}=10^{-5}$~J/m$^2$ (see $z=0$), and 
           $W^{(2)}_\text{wall}=0.1\times10^{-5}$~J/m$^2$ (see $z=D$), $K=9\times 10^{-12}$ N \cite{footnote3} (the values of the remaining parameters are the same as in Fig.\ \ref{fig:Hyst}).}
   \label{fig:TwoWallsVectors}
\end{figure}

\begin{figure*}[t!]
   \centering
   \includegraphics[scale=1]{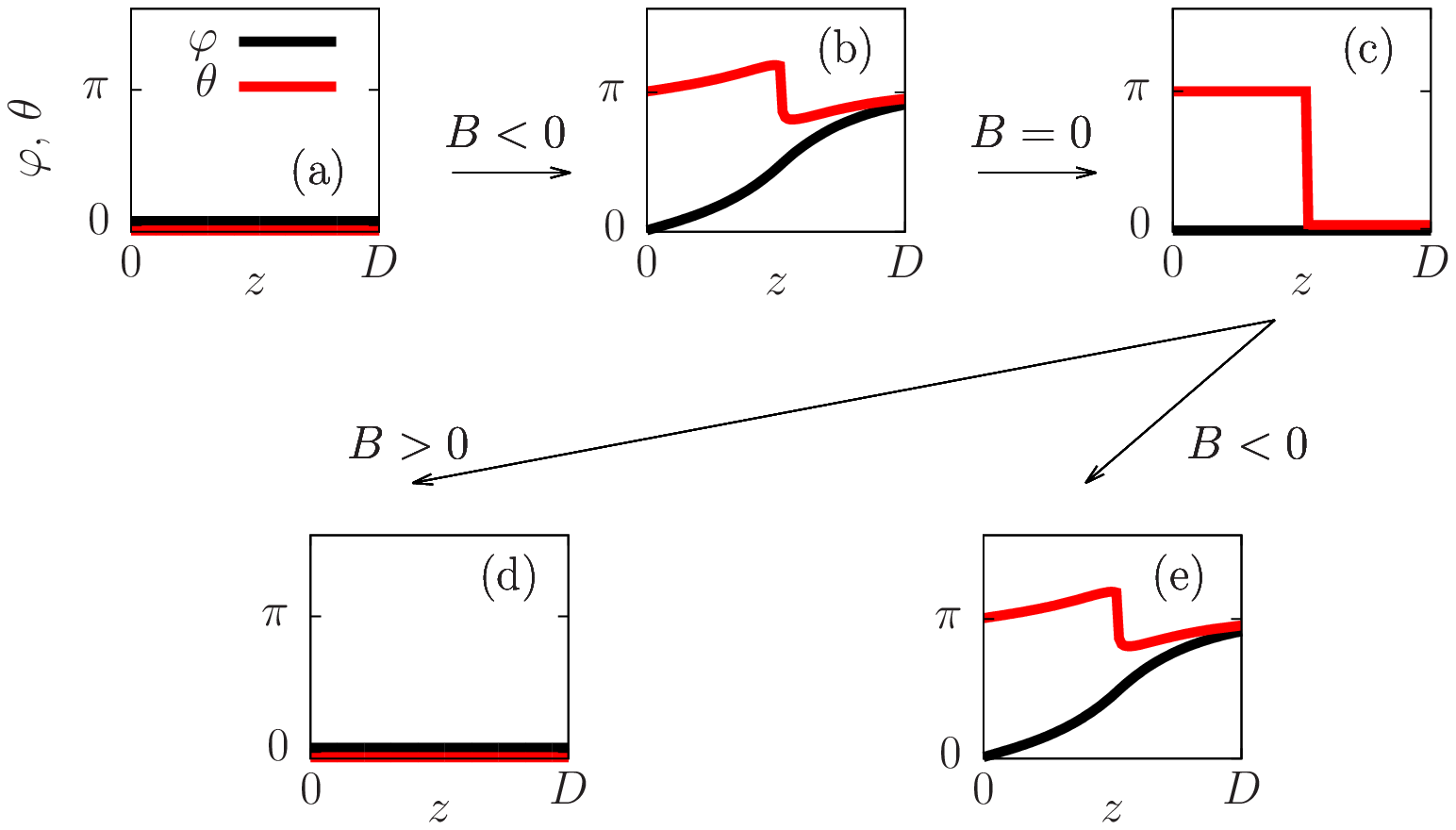}
   \caption{Orientational profiles $\varphi(z)$ (nematic director, black line) and $\theta(z)$ 
           (magnetization, red line) in a ferronematic cell with walls of different 
           anchoring strength, as discussed in Sec.~\ref{subsec:TwoWalls}. 
           The initial state (a) is a uniformly magnetized 
           ferronematic slab, i.e., the director field ($\varphi=0$, black line) and
           the magnetization field ($\theta=0$, red line) are uniform. 
           Upon application of an external magnetic field $\mathbf{B}=B\mathbf{e_x}$
           in the direction opposite to the initial magnetization direction (i.e., 
           $B<0$) transfers the sample into the disturbed state (b) 
           (compare Fig.~\ref{fig:TwoWalls}). 
           After suddenly switching off the external field (i.e., for $B=0$) the
           system relaxes into the state (c) in which the nematic director is uniform;
           that part of the sample, which is close to the wall with strong anchoring, retains its 
           magnetization direction $\theta\approx \pi$ whereas the magnetization near the
           wall with weak anchoring follows the relaxation of the nematic director and attains
           $\theta=0$.
           Thus having two halves of the sample being magnetized in opposite directions 
           yields zero overall sample magnetization, i.e., $\mathcal{M}\ll 1$. 
           This configuration offers two options: 
           (i) The application of an external magnetic field in the direction of the 
           initial magnetization (i.e., $B>0$) returns the sample to the
           initial, uniform state (d)=(a) with magnetization $\mathcal{M}=1$. 
           (ii) The application of an external magnetic field in the direction opposite
           to the initial magnetization (i.e., $B<0$) produces the disturbed state 
           (e)=(b). 
           The fact that the states (d) and (a) are identical allows one to cycle through
           the states (a), (b), and (c) by applying the external magnetic field 
           $\mathbf{B}$ in suitable directions.
           The width of the interface between two domains in (c) is not larger than the numerical grid discretization, i.e., less than $D/100$.
           The values of the parameters are the same as in Fig.\ \ref{fig:TwoWalls}.}
   \label{fig:Storage}
\end{figure*}

In the middle ($z\approx 0.55D$) of the sample the magnetization field exhibits an interface between two halves of the slab (see Fig.~\ref{fig:TwoWallsVectors}). 
The orientation of the magnetization within the two halves differs only in how the magnetization approaches the value $\theta=\pi$ in the vicinity of the center of the slab. 
This behavior of the magnetization field profile is caused by the necessity to be 
compatible with the nematic director profile in the center region. 
This means that the rotation of the nematic director in the interior of the slab forces the magnetization direction to reach its value $\theta=\pi$ at $z\approx D/2$ either from $\theta>\pi$ at $z\lesssim D/2$ or from $\theta<\pi$ at $z\gtrsim D/2$ (see the red curve in Fig.\ \ref{fig:TwoWalls}). 
In the situation of Fig.~\ref{fig:TwoWalls}, upon switching off the external magnetic field, 
we found that the system relaxes into a state with a uniform nematic director field and two domains with the magnetization pointing into opposite directions (see Fig.~\ref{fig:Storage}(c)) \cite{footnote3}.
The position of the interface between these two domains depends on the position of the interface plane formed while the magnetic field was still on \cite{footnote4}. 
Application of the external magnetic field to the two-domain configuration opens up two possibilities: (i) If the external magnetic field is applied in the same direction as the field used to create the two-domain sample, the resulting state is identical to the one in Fig.~\ref{fig:TwoWalls}. (ii) If, on the other hand, the magnetic field is applied opposite to the direction of the magnetic field used to create the two-domain sample, one of the domains (i.e., the one the magnetization of which is opposite to the external field) switches. This yields a uniform sample both in terms of the nematic director and the magnetization field, thus returning the system to its initial state. 
These steps are summarized in Fig.~\ref{fig:Storage}. 
Note that the states shown in Figs.~\ref{fig:Storage} (a) and \ref{fig:Storage} (d) are identical. 
Also note that the state depicted in Fig.~\ref{fig:Storage} (a) exhibits saturated magnetization $\mathcal{M}=1$ (Eq.\ \eqref{eq:PrettyM}) whereas the state depicted on Fig.~\ref{fig:Storage} (c) exhibits $\mathcal{M}\ll 1$. 
Since one can restore the initial state (see Fig.~\ref{fig:Storage} (a)) from the two-domain state (see Fig.~\ref{fig:Storage}(c)) by applying an external magnetic field $B>0$, one is able to cycle through three states (see Figs.~\ref{fig:Storage} (a), (b), and (c)). 
Accordingly, this ferronematic cell with two walls of different anchoring strength can be put in either of two states (i.e., magnetized or demagnetized) by using an external magnetic field of suitable direction. This opens up the possibility, e.g., to use an array of such cells for storage of binary information with a "bit" being represented by the state of the cell (magnetized/demagnetized) or as a spatially resolving magnetic field detector with memory function.

\begin{figure}[t!]
   \centering
   \includegraphics[scale=1]{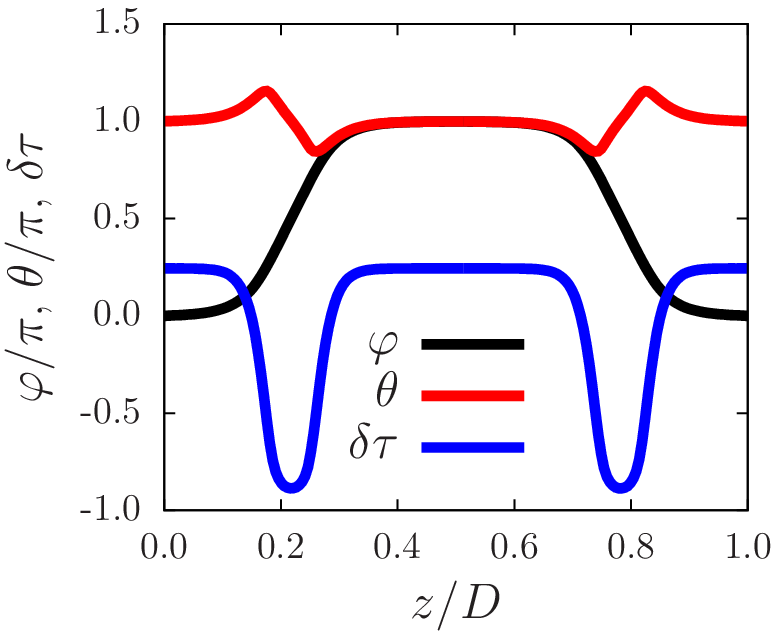}
   \caption{Results of the numerical minimization of the functional in Eq.~\eqref{eq:SlabF} with segregation effects included (see Eqs.~\eqref{eq:Split} and \eqref{eq:FSeg}). 
       The sample thickness is $D=120\,\mathrm{\mu m}$, the wall anchoring is $W_\text{wall}=3.4\times10^{-5}$~J/m$^2$ for both walls, the coupling constant is $\gamma=240$, the external magnetic field is $B=4$~mT, and $a\approx 3.1\times 10^{-4}$~N/A$^2$. (The values of the remaining parameters are the same as in Fig.\ \ref{fig:Hyst}.) 
       Due to the large thickness $D$ of the cell the switching mechanism I is now combined with the switching mechanism II even in the case that both walls provide strong anchoring. 
       The particles are expelled (see the blue solid line) from the regions with strong gradients of the nematic director field $\varphi$ (black solid line). 
       The depletion layers (i.e., the minima of the blue solid line) are separated from the walls due to the soft anchoring between the magnetization and the nematic director field. This differs from the situation described in Ref.~\cite{1970_Brochard} in which the depletion layer is located in close vicinity of the wall.}
   \label{fig:Segregation}
\end{figure}

\section{\label{sec:Segregation}Segregation effects}
It was pointed out by Brochard and de Gennes \cite{1970_Brochard} that anisotropic magnetic colloids tend to move away from regions of the NLC where distortions of the director field prevent them from minimizing their free energy in the external magnetic field. 
The segregation parameter defined as (see Ref.~\cite{1970_Brochard})
\begin{equation}
  s:=\beta mB
\end{equation}
is greater than unity already for $B\gtrsim 1.4$ mT and therefore one can expect segregation to occur for external fields stronger than $1.4$ mT. 
In the following we investigate the impact of segregation on the switching mechanisms I and II. 

The present theoretical approach (see Eqs.~\eqref{eq:FN} and \eqref{eq:SlabF}) includes the possibility of segregation to occur through the dependence of the magnetization field $\mathbf{M}$ on the spatial coordinate $z$. 
In particular, we are interested in spatial inhomogeneities of the \textit{absolute value} $|\mathbf{M}(z)|$ of the magnetization vector. 
It is convenient to introduce the dimensionless quantity $\tau (z) := |\mathbf{M}(z)|/(m\rho_\text{iso})$. 
So far all our results have been obtained in the limit $\tau(z) = \text{const} =1$. 
If $\tau (z) \neq 1$, that part of the free energy density, which depends on it (see Eq.~\eqref{eq:FN} and Fig.\ \ref{fig:Slab}), is given by
\begin{equation}
  \begin{aligned}
    \label{eq:FNWithS}
    \beta &f(\theta(z), \varphi(z), \tau(z))=\\
    &\beta (m\rho_\text{iso})^2\tau(z)^2 \Big(\frac{a}{2}-\frac{1}{2}\gamma\mu_0 \cos(\theta(z)-\varphi(z))^2\Big)\\
    &-\beta m\rho_\text{iso}B\tau(z)\cos(\theta(z)-\psi).
  \end{aligned}
\end{equation}
Concerning the segregation effects the value of $a$ matters. 
According to Ref.\ \cite{2018_Zarubin} the value of $\gamma=240$ implies $c\approx 0.035$ and therefore $a\approx 10\times k_\text{B}T/(m^2\rho_\text{iso})\approx 3.1\times 10^{-4}$ N/A$^2$ (concerning the definition of $a$ in terms of $c$ see Ref.\ \cite{2018_Zarubin}).
Since the sample always contains a fixed number of magnetic colloids (i.e., neglecting aggregation) the field $\tau(z)$ is subject to the constraint (see Appendix \ref{sec:Appendix})
\begin{equation}
  \label{eq:TauCons}
\frac{1}{D}\int_0^D\!\!\dd z\,\,\tau(z)=1.
\end{equation}
It is convenient to consider deviations $\delta\tau(z)$ from the homogeneous case, i.e.,
\begin{equation}
  \tau(z)=1+\delta\tau(z),
  \label{eq:deltataudef}
\end{equation}
which allows one to rewrite the free energy density in Eq.~\eqref{eq:FNWithS} as the sum of the free energy density evaluated for $\tau(z) = \text{const} =1$ and the contribution due to segregation:
\begin{align}
  \label{eq:Split}
  \beta &f(\theta(z), \varphi(z), \tau(z))=\nonumber\\
  &\beta f(\theta(z),\varphi(z), \tau(z)=1)+\beta f_\text{seg}(\theta(z), \varphi(z), \tau(z)),
\end{align}
where $f_\text{seg}(\theta(z), \varphi(z), \tau(z))$ is defined as
\begin{equation}
  \label{eq:FSeg}
  \begin{aligned}
    \beta &f_\text{seg}(\theta(z), \varphi(z), \tau(z)):=\\
    &\beta (m\rho_\text{iso})^2\Big(a-\gamma\mu_0 \cos(\theta(z)-\varphi(z))^2\\
    &-\beta m\rho_\text{iso}B\cos(\theta(z)-\psi)\Big)\delta\tau(z)\\
    &+\beta (m\rho_\text{iso})^2 \Big(\frac{a}{2}-\frac{1}{2}\gamma\mu_0 \cos(\theta(z)-\varphi(z))^2\Big)\delta\tau(z)^2.
  \end{aligned}
\end{equation}
The constraint in Eq.~\eqref{eq:TauCons} turns into
\begin{equation}
  \label{eq:DTauCons}
  \int_0^D\!\!\dd z\,\,\delta\tau(z)=0.
\end{equation}
We express $\delta\tau(z)$ in terms of a Fourier series:
\begin{equation}
  \delta\tau(z)=\frac{a_0}{2}+\sum_{n=1}^\infty \Big[a_n\cos\Big(\frac{2\pi nz}{D}\Big)+b_n\sin\Big(\frac{2\pi nz}{D}\Big)\Big].
\end{equation}
Equation~\eqref{eq:DTauCons} implies $a_0=0$. 
The functional in Eq.~\eqref{eq:SlabF} is minimized with respect to the fields $\theta(z)$ and $\varphi(z)$ and the coefficients $a_n$ and $b_n$, $n\in \{1,2,...,N\}$. 
The number of coefficients $N$ to be taken into account has to be chosen. 
It is reasonable to set the minimum wavelength in the Fourier series to be larger than the colloid diameter $d\sim 100$ nm. 
Therefore, $N$ has to be smaller than $N_\text{max}=[D/d\,]$ where $[x]$ denotes the integer part of $x$. 
For a slab of thickness $D=20$~$\mu$m one has $N_\text{max}=200$.

It turned out that for slab thicknesses $D<60\,\mathrm{\mu m}$ the equilibrium profile
$\tau(z)$ obtained from Eq.~\eqref{eq:FNWithS} takes negative values, i.e., 
$\delta\tau(z)<-1$, which contradicts its physical meaning $\tau(z)\sim|\mathbf{M}(z)|
\geq0$.
This behavior is related to the absence of contributions in $\tau(z)$ beyond 
quadratic order.
However, for slab thicknesses $D\geq 60$ $\mu$m the algorithm does provide the profiles $\theta(z)$ and $\varphi(z)$ together with a physically reasonable segregation profile
$\delta\tau(z)$. 
Figure \ref{fig:Segregation} shows the calculated profiles for $D= 120$~$\mu$m, equal walls with strong anchoring $W_\text{wall}=3.4\times10^{-5}$~J/m$^2$, coupling constant $\gamma=240$, and external magnetic field $B=4$~mT. 
It is evident that switching mechanism I is observed even in the presence of segregation effects. 
The density of magnetic colloids is largely reduced (with $\delta\tau(z)$ close to -1)
in the regions of nonzero gradient of the director profile. 
An important difference to the case of infinitely strong coupling of the colloids to the liquid crystal (see Ref.~\cite{1970_Brochard}) is the \textit{depletion layer} being shifted away from the walls towards the interior of the sample. 
The functional form of the profiles $\varphi(z)$ and $\theta(z)$ obtained for asymmetric pairs of walls with strong and weak anchoring (see Fig.~\ref{fig:TwoWalls} in Sec.~\ref{subsec:TwoWalls}) is found at both walls for sufficiently thick slabs. 
We have performed a series of calculations for different slab thicknesses $D$ in order
to determine for which thickness the pure switching mechanism I turns into a
combination of mechanisms I and II (see Fig.~\ref{fig:Segregation}); we have found $D=95\pm 5$~$\mu$m. 
This observation opens the possibility to manipulate the sample magnetization in a manner similar to that described in Sec.~\ref{subsec:TwoWalls} but without the need to use a second, weakly anchored wall. 
However, the state of nonzero net magnetization is not necessarily saturated,
but it might exhibit $\mathcal{M} < 1$. 
In order to have nonetheless a state with the magnetization $\mathcal{M}\approx 1$, one would need to adjust the system parameters (e.g., wall anchoring, elastic constant of the NLC etc.) such that each region of switched magnetization in the vicinity of the walls (see Fig.~\ref{fig:Segregation}) takes up $\approx 25\%$ of the slab thickness.

We have also performed a calculation for the case of two walls with equally weak anchoring. 
We found that segregation did not have a qualitative impact in that case either. 
Therefore we expect that the results of Secs.\ \ref{subsec:I} and \ref{subsec:II} are not affected qualitatively by segregation.


\section{\label{sec:Discussion}Summary and conclusions}

In this analysis we have studied theoretically a ferronematic confined
between two planar, parallel walls which impose an easy axis on the NLC director field. 
Inspired by the experimental studies reported in Refs. \cite{2013_Mertelj, 2014_Mertelj}, the system is subjected to an external magnetic field. 
The ferronematic is an anisotropic polar fluid and thus the system is characterized by the
relative directions of the NLC director, the easy axes due to the walls, the magnetization,
and the external magnetic field. 
We have considered the situation in which the ferronematic is initially prepared with a uniform
magnetization along the easy axis of the NLC.
Subsequently an external magnetic field is applied in the direction opposite to the magnetization. 
This choice of the geometry reduces the theoretical description to an effectively one-dimensional one. 
The experiments reported in Refs.~\cite{2013_Mertelj, 2014_Mertelj} showed that for such a setup 
there exists a critical external magnetic field  $B_\text{cr}>0$ such that for magnetic
field strengths $|\mathbf{B}|<B_\text{cr}$ the sample remains unperturbed. 
The authors of Ref.~\cite{2013_Mertelj} also provided the expression in Eq.~\eqref{eq:BCr}
for the critical magnetic field strength in terms of the coupling $\gamma$ between the
magnetization and the nematic director field. 
This critical magnetic field strength, which increases upon increasing $\gamma$, has been
determined as that magnetic field strength for which the relaxation rate of long-wavelength
fluctuations of the nematic director field vanish.

Here we study the system by numerical minimization of the corresponding 
free energy functional in Eq.~\eqref{eq:SlabF}. 
The numerical minimization is performed by using the Fletcher-Reeves-Polak-Ribiere general
function minimization algorithm. 
It is obvious, that the global minimum of the free energy functional in Eq.~\eqref{eq:SlabF} before the
external magnetic field has been applied is the initial state of the ferronematic
being uniformly magnetized along the easy axis. 
Once the magnetic field is applied in the direction opposite to the initial magnetization,
this state becomes only a local (metastable) minimum. 
The new global minimum is the ferronematic magnetized in the direction of the field. 
By means of a conjugate gradient algorithm one is able to search for the local minimum of 
the free energy and therefore to identify metastable states of the system. 
This is particularly useful in the present context, because this way one can investigate possible intermediate orientation profiles between the initial, now metastable, state with a uniform 
magnetization in the positive $x$-direction, i.e., opposite to the magnetic field $\mathbf{B}=B\mathbf{e}_x$ pointing into the negative $x$-direction, i.e., $B<0$, and the final stable state
with the magnetization in the direction along the magnetic field, i.e., in negative $x$-direction ($B<0$).

Figure \ref{fig:Hyst} shows the dependence of the dimensionless magnetization $\mathcal{M}$
(Eq.~\eqref{eq:PrettyM}) of the metastable state described above on the strength of the
external magnetic field for particular values of the coupling constant $\gamma$ and of the anchoring strength $W_\text{wall}$ at a wall. 
One observes hysteresis of the magnetization for which a critical magnetic field strength
$B_\text{cr}$ can be identified as the one for which significant deviations from the saturation
magnetization $\mathcal{M}=1$ occur. 
One can distinguish several, qualitatively different, intermediate states (red circles in 
Fig.~\ref{fig:Hyst}). 
First, for a magnetic field $\mathbf{B}=B\mathbf{e}_x$, with component $B$ in the direction of the initial
magnetization ($B>0$), the sample remains practically unperturbed for $B > B_\text{cr}$ ($B_\text{cr}<0$), i.e., for
the magnetic field either along the initial magnetization ($B>0$) or sufficiently weak in the 
direction opposite to the initial magnetization ($B<0$) (see Fig.~\ref{fig:MetaS}(1)). 
Upon decreasing the component $B$ of the magnetic field $\mathbf{B}=B\mathbf{e}_x$ further (i.e., making it less positive or more negative), the torque imposed
on the NLC by the walls is no longer able to keep the ferronematic in the initial unperturbed
state and thus the profiles become perturbed (see Fig.~\ref{fig:MetaS}(2)). 
These states correspond to the brightening of the sample when viewed via crossed polarizers \cite{2013_Mertelj}. 
If one decreases the magnetic field component $B$ even further (i.e., making $B$ even more negative), one encounters the interesting metastable 
state shown in Fig.~\ref{fig:MetaS}(3).
In this state, near each wall a layer is formed within which
the nematic director is close to the easy axis and the magnetization has inverted its 
direction, pointing along the external magnetic field (i.e., in negative $x$-direction). 
This flipping of the magnetization is energetically favorable for a sufficiently large
strength $|\mathbf{B}|=|B|$ of the magnetic field pointing in the direction opposite to the initial
magnetization (i.e., pointing into the negative $x$-direction), because the contribution to the free energy (Eq.~\eqref{eq:FN}) of the 
coupling between the magnetization and the nematic director is invariant upon inversion of
the magnetization, $\mathbf{M} \mapsto -\mathbf{M}$, but the contribution of the coupling
between the magnetization and the external magnetic field is not. 
When the external magnetic field becomes even stronger (i.e., $B$ becomes even more negative and $|B|$ even larger), the regions of flipped magnetization
expand into the interior of the system, eventually giving rise to the whole sample (except for
thin layers in the very vicinity of the walls) being magnetized along the magnetic field. 
The qualitatively different scenario, which we refer to as switching mechanism II, occurs if the wall anchoring $W_\text{wall}$ is too weak to prevent the nematic director field from following the rotating magnetization. 
In accordance to Figs.\ \ref{fig:MetaS}(3) and \ref{fig:MetaS}(4), in this scenario the final stages of the switching are not realized. 
Figure \ref{fig:MetaSII} illustrates how the final stages (3) and (4) of the switching mechanism II are realized, according to which the magnetization \textit{and} the director field rotate in parallel.

Figure \ref{fig:Switch} shows whether certain combinations of wall anchoring strengths $W_\text{wall}^{(1)}=W_\text{wall}^{(2)}=W_\text{wall}$ as well as of the coupling constant
$\gamma$ lead to switching mechanism I or II. 
According to the map in Fig.\ \ref{fig:Switch} the switching mechanism I is the dominant one in the experiments described in Refs.\ \cite{2013_Mertelj,2014_Mertelj} ($W_\text{wall}^\text{exp}\approx 3.4\times 10^{-5}$~J/m$^2$).

The dependences of the critical magnetic field strength $B_\text{cr}$ on the
coupling constant $\gamma$ and on the (equal) wall anchoring strength $W_\text{wall}$ are presented in
Figs.~\ref{fig:Phase} and \ref{fig:Cuts}. 
On one hand, $B_\text{cr}$ increases as function of $W_\text{wall}$ and, on the other
hand, it also increases as function of $\gamma$, which is consistent with the results of
Ref.~\cite{2013_Mertelj}. 
In Fig.~\ref{fig:Cuts}(b) a comparison of the critical magnetic field strength 
$B_\text{cr}$ as defined here with that introduced in Ref.~\cite{2013_Mertelj}
(see Eq.~\eqref{eq:BCr}) shows good agreement, although the two expressions involve 
different properties of the ferronematic.

Within a recently developed theory of ferronematics \cite{2018_Zarubin} one can relate the
coupling coefficient $\gamma$ to the microscopic coupling $c$ (see below Eq.\ \eqref{eq:FN}) which depends on the size of the colloids in the suspension. 
Figure~\ref{fig:Cuts}(c) shows the dependence of $B_\text{cr}$ on the microscopic coupling
$c$ for a particular value of the wall anchoring strength $W_\text{wall}$. 
This allows one to vary the critical magnetic field by tuning the mean value of the size distribution of the colloids participating in the ferronematic.

Combining two walls with different anchoring strengths allows one to design a sample
such that its switching mechanism is a superposition of type I and type II. 
The resulting nematic director and magnetization field profiles (see 
Figs.~\ref{fig:TwoWalls} and \ref{fig:TwoWallsVectors}) are obtained by applying an external magnetic field. 
In turn, switching off this field divides the sample into two domains with opposite magnetizations (see Fig.~\ref{fig:Storage} (c)), rendering a sample with zero net magnetization. 
The initial state (i.e., the magnetized slab) can be restored by applying an external magnetic field of suitable direction to the two-domain sample. 
This cycle can be repeated arbitrarily, thus facilitating the switching between two states (magnetized/demagnetized slab) by using a uniform magnetic field only. 
This opens application perspectives such as storage of information and magnetic fields detection.

Similar controllable magnetic slabs can be constructed by using two walls with equally strong anchoring for samples of larger thickness ($D\geq 95$ $\mu$m for $\gamma=240$, $K=3.5\times 10^{-12}$ N, and $W_\text{wall}=3.4\times 10^{-5}$ J/m$^2$). 
We have found segregation to be \textit{quantitatively} different from the case of walls with infinitely strong anchoring and of infinitely strong coupling of the colloids to the NLC (see Ref.\ \cite{1970_Brochard}). 
However, segregation effects do not affect the switching mechanisms qualitatively.


\section*{Conflicts of interest}
There are no conflicts to declare.

\appendix
\section{\label{sec:Appendix}Constraint of the field $\tau(z)$}

In this Appendix we derive Eq.\ \eqref{eq:TauCons}. 

The spatially varying magnetization field $\mathbf{M}(z)$ was defined in Ref.~\cite{2018_Zarubin} as
\begin{equation}
  \mathbf{M}(z)=\int\!\!\dd^2\omega\,\, m\bs{\omega}\,\rho(z,\bs{\omega}),
\end{equation}
where $m$ is the magnitude and $\bs{\omega}$ the direction of the magnetic moment of a single colloid and $\rho(z,\bs{\omega})$ is the number density of colloids in a layer around point $z$ and oriented in direction $\bs{\omega}$. 
We \textit{assume} that in a small layer around a given point $z$ all individual magnetic moments point in one direction, i.e., the direction of $\mathbf{M}(z)$:
\begin{equation}
\rho(z,\bs{\omega})=g(z)\delta(\bs{\omega}-\bs{\omega}_0(z)),
\end{equation}
with $\bs{\omega}_0(z):=\mathbf{M}(z)/|\mathbf{M}(z)|$ and $g(z)$ is the number density of colloids at point $z$ regardless of their orientation. 
From the definition of $\rho_\text{iso}$ it follows 
\begin{align}
   \rho_\text{iso} = 
   \frac{1}{D}\int_0^D\!\!\dd z\int\!\!\dd^2\omega\,\,\rho(z,\bs{\omega}) =
   \frac{1}{D}\int_0^D\!\!\dd z                   \,\,g(z).
   \label{eq:rhoisointg}
\end{align}
Noting that
\begin{equation}
  \label{eq:AbsM}
  |\mathbf{M}(z)|=mg(z)
\end{equation}
and defining $\tau (z) := |\mathbf{M}(z)|/(m\rho_\text{iso}) = g(z)/\rho_\text{iso}$,
Eq.~\eqref{eq:rhoisointg} can be written in the form of Eq.~\eqref{eq:TauCons}:
\begin{equation}
  \label{eq:IntG}
  1 = 
  \frac{1}{D}\int_0^D\dd z\,\,\frac{g(z)}{\rho_\text{iso}} = 
  \frac{1}{D}\int_0^D\dd z\,\,\tau(z).
\end{equation}
Using the definition $\delta\tau(z) := \tau(z) - 1$ (see Eq.~\eqref{eq:deltataudef}) this is
equivalent to (see Eq.~\eqref{eq:DTauCons})
\begin{equation}
  \int_0^D\!\!\dd z\,\,\delta\tau(z)=0.
\end{equation}


\begin{thebibliography}{4}
\bibitem{footnote1}
Segregation effects can be neglected if the segregation parameter $\kappa:=
D/l$ is much larger than unity \cite{2014_Zakhlevnykh}, where 
$l:=\sqrt{K/(k_\text{B}T\rho_\text{iso})}$ is the segregation length \cite{1970_Brochard}; for $K$ see Eq.\ \eqref{eq:NLCEl}. 
In the present study one has $\kappa\approx 7$.
\bibitem{footnote5}
The positive constant $a$ in Eq.\ \eqref{eq:FN} is unimportant as long as $|\mathbf{M}|=m\rho_\text{iso}=\text{const}$. It becomes relevant otherwise, i.e., if segregation is taken into account (see Sec.\ \ref{sec:Segregation}).
\bibitem{footnote3}
The two-domain state required us to tune the elastic constant $K$, because for certain values of $K$ the minimization algorithm produced throughout the sample a perturbed director profile for $B=0$ as a favorable state. 
In order to make the corresponding elastic distortions more costly, we increased the value of $K$ to $9\times 10^{-12}$ N compared to $K=3.5\times 10^{-12}$ N used in Secs.\ \ref{subsec:I}, \ref{subsec:II}, and \ref{sec:Segregation}.
\bibitem{footnote4}
The position of the interface depends on the external field \textit{strength} which should be chosen such that the interface is located approximately in the middle of the slab. We used the values $B=12$~mT together with $\gamma\approx 240$, $W_\text{wall}^{(1)}=1\times 10^{-5}$ J/m$^2$, $W_\text{wall}^{(2)}=0.1\times 10^{-5}$ J/m$^2$, and $K=9\times 10^{-12}$ N.
\end{thebibliography}
\end{document}